\documentclass[10pt]{article}

\usepackage[letterpaper]{geometry}
\usepackage{hicss}
\usepackage{times}
\usepackage[none]{hyphenat}
\usepackage{url}
\usepackage{latexsym}
\usepackage{graphicx}
\graphicspath{{images/}}

\usepackage{amsmath,amssymb,amsbsy,amsfonts,amsthm}
\usepackage{enumitem}
\usepackage{tikz}
\usetikzlibrary{trees}
\tikzstyle{level 1}=[level distance=3.5cm, sibling distance=3.5cm]
\tikzstyle{level 2}=[level distance=3.5cm, sibling distance=2cm]

\tikzstyle{bag} = [text width=5em, text centered]
\tikzstyle{end} = [circle, minimum width=3pt,fill, inner sep=0pt]

\title{Robustness in Nonorthogonal Multiple Access 5G Networks
\thanks{\hspace{.2cm}This paper has been accepted for the upcoming 55th Hawaii International Conference on System Sciences (HICSS-55).}
}

\author{Benjamin Pimentel \\
  Naval Postgraduate School \\
  {\underline{ benjamin.pimentel@nps.edu}} \\\And
  Alex Bordetsky \\
  Naval Postgraduate School \\
  {\underline{ abordets@nps.edu} }\\\And 
  Ralucca Gera \\
  Naval Postgraduate School \\
  {\underline{ rgera@nps.edu}} \\}

\date{}

\begin{document}
\maketitle
\begin{abstract}
The diversity of fifth generation (5G) network use cases, multiple access technologies, and network deployments requires measures of network robustness that complement throughput-centric error rates. In this paper, we investigate robustness in nonorthogonal multiple access (NOMA) 5G networks through temporal network theory. We develop a graph model and analytical framework to characterize time-varying network connectedness as a function of NOMA overloading. We extend our analysis to derive lower bounds and probability expressions for the number of medium access control frames required to achieve pairwise connectivity between all network devices. We support our analytical results through simulation.
\end{abstract}

\section{Introduction} \label{intro}
The pursuit of 5G wireless network performance objectives in \cite{series2015imt} has led to many advances in radio access network technology including nonorthogonal multiple access (NOMA) \cite{dahlman20185g,ahmadi20195g}. NOMA is a family of multiple access approaches that assign multiple network devices (ND) to transmit or receive radio frequency (RF) signals using nonorthogonal time-frequency resources, or resource blocks (RB). Unlike orthogonal multiple access (OMA) schemes that separate ND transmissions in frequency, time, code, space, or some combination thereof, NOMA allows simultaneous use of RBs, and employs a variety of multiuser detection (MUD) methods to separate the signals \cite{chen2018toward}. NOMA methods are broadly categorized as power-domain or code-domain. Power-domain approaches differentiate between NDs using a combination of superposition coding and successive interference cancellation while code-domain schemes achieve MUD through sparse or nonorthogonal low cross-correlation spreading sequences \cite{islam2016power,dai2018survey,vaezi2019multiple,shahab2020grant}. 

A ratio of NDs to RBs that is greater than one is called \emph{overloading}. Through overloading, NOMA has the potential to support increased aggregate throughput, improved spectral efficiency, lower latency, and higher connection densities than OMA \cite{dai2018survey,vaezi2019multiple,shahab2020grant,mostafa2019connection}. Thus, the impact of overloading on network robustness is an important consideration for 5G NOMA network design. 

While NOMA is not limited to 5G networks, we focus on 5G New Radio (NR) given the body of 5G-related NOMA research \cite{islam2016power,dai2018survey,vaezi2019multiple,ding2015impact,yuan20205g} and ongoing discussion of NOMA in the 3rd Generation Partnership Project \cite{3gpp.38.812}. We find specific motivation for the study of NOMA robustness in the context of 5G NR aerial access networks (AAN) \cite{3gpp.38.811}. AANs can provide 5G service for underserved areas and disaster recovery/public safety via unmanned aerial vehicle (UAV) base stations \cite{mozaffari2019tutorial}. Dense network deployments are less likely in these scenarios making the sub-6 GHz range particularly important. Though much emphasis is placed on massive multiple input multiple output (mMIMO) to achieve 5G performance \cite{dahlman20185g}, the spatial multiplexing gains of mMIMO are less feasible in sub-6 GHz AANs due to a combination of increased channel correlation for LOS links \cite{ nasir2019uav} and the size, weight, and power requirements of a mMIMO array \cite{ bjornson2019massive}. Thus, NOMA is a more prominent candidate to support 5G performance objectives in sub-6 GHz AANs.  

In this paper, we investigate the relationship between NOMA overloading and network robustness through temporal graph components. We develop a graph model and analytical framework to characterize temporal component membership. We extend our analysis to derive lower bounds and probability expressions for the number of medium access control (MAC) frames required to achieve strong temporal connectedness between all NDs. These measures of network robustness are relevant for real-time Internet of Things \cite{park2020centralized} and distributed computing applications \cite{nedic2018network,shi2020communication} with small message lengths in which RB allocation for timely message exchange is an important factor in application performance.

The remainder of this paper is organized as follows. Section \ref{relwork} discusses related work. Section \ref{temp} introduces the concepts of temporal connectedness and components. Section \ref{NOMAmodel} introduces our NOMA graph model. Section \ref{framework} describes our stochastic temporal component framework followed by individual and joint temporal component analysis and Sections \ref{indprobtempmem}-\ref{jointprobtempmem}. We present our affine graph analysis and simulation results in Section \ref{affineanalysis}, and conclude the paper in Section \ref{conc}. 

\section{Related work} \label{relwork}
The robustness of wireless networks is often approached through stochastic geometry (SG) analysis due to the interference-limited nature of OMA \cite{haenggi2009stochastic,win2009mathematical,elsawy2016modeling}. SG is a well-studied approach that provides robustness metrics such as outage probability and bit error probability through the spatial averaging of random point processes, and has been applied for NOMA analysis \cite{ali2018downlink,yue2018unified}. Recent SG analysis also includes a temporal aspect through queuing theory to consider maximum traffic density and age of information \cite{chisci2019uncoordinated,emara2020spatiotemporal}. However, the concept of overloading reduces the primacy of interference-based analysis since NDs in NOMA networks interfere by design. Therefore, investigating NOMA network robustness through temporal graph component metrics is an appropriate complement to SG analysis.

Much of the NOMA literature investigates methods to achieve 5G performance objectives under practical constraints \cite{chen2018toward,dai2018survey,vaezi2019multiple,ding2015impact,ding2014performance,shirvanimoghaddam2017fundamental,makki2020survey}. Within this body of work, several researchers have considered overloading through standard physical and network layer measures of robustness. The link level simulation results presented in \cite{Huawei_HiSilicon} compare uplink NOMA to OFDMA by assessing the effect of variable overloading on block error rate at multiple signal-to-noise ratios. Stability conditions for uplink code-domain NOMA are derived in \cite{shirvanimoghaddam2017fundamental} where stability is defined in terms of the queue size of NDs attempting to transmit in each time slot. Similarly, overloading stability in different downlink power-domain NOMA scenarios is examined in \cite{huang2019optimal} where stability is defined as the power allocation among NDs that is required for successful signal differentiation. Our contribution provides a NOMA graph model and analytical framework that measures the effect of variable overloading on temporal connectedness between NDs.

\section{Temporal connectedness and components} \label{temp}
Network connectedness is a common measure of function and robustness in static networks \cite{bianconi2018multilayer,newman2018networks}. Measures of connectedness for static networks are extended to temporal (time-varying) networks in ~\cite{nicosia2012components,nicosia2013graph}. Unlike a static network, two nodes in a temporal network may only be connected for some finite duration of time, called a contact. A time-ordered sequence of contacts that allows two nodes to reach each other without traversing an edge more than once is a temporal path. Thus, node $i$ is \textit{temporally connected} to node $j$ if there exists a temporal path from $i$ to $j$. 

Temporal connectedness is not a symmetric relation (i.e., a temporal path from node $i$ to $j$ does not imply a temporal path from node $j$ to $i$), giving rise to definitions of strong and weak connectedness similar to directed static networks. Two nodes $i$ and $j$ of a time-varying graph are strongly connected if $i$ is temporally connected to $j$, and also $j$ is temporally connected to $i$. Similarly, two nodes $i$ and $j$ of a time-varying graph are weakly connected if $i$ is temporally connected to $j$, and $j$ is also temporally connected to $i$ in the underlying undirected time-varying graph. This sets up the following temporal component definitions~\cite{nicosia2012components,nicosia2013graph}:
\begin{itemize}[noitemsep,topsep=0pt,parsep=0pt,partopsep=0pt]
\item \textit{Temporal Strongly Connected Components} (TSCC): A set of nodes of a time-varying graph in which each node of the set is temporally strongly connected to all other nodes in the set. 
\item \textit{Temporal Weakly Connected Components} (TWCC): A set of nodes of a time-varying graph in which each node of the set is temporally weakly connected to all other nodes in the set.

\end{itemize}

Finally, the \emph{affine graph} of a time-varying network is the underlying undirected static graph associated with a specified time interval such that two nodes $i$ and $j$ are connected in the affine graph only if they are strongly connected in the time-varying graph \cite{nicosia2012components,nicosia2013graph}. 

\section{NOMA graph model} \label{NOMAmodel}

\begin{figure}[!ht]
    \centering
	\includegraphics[scale=0.6]{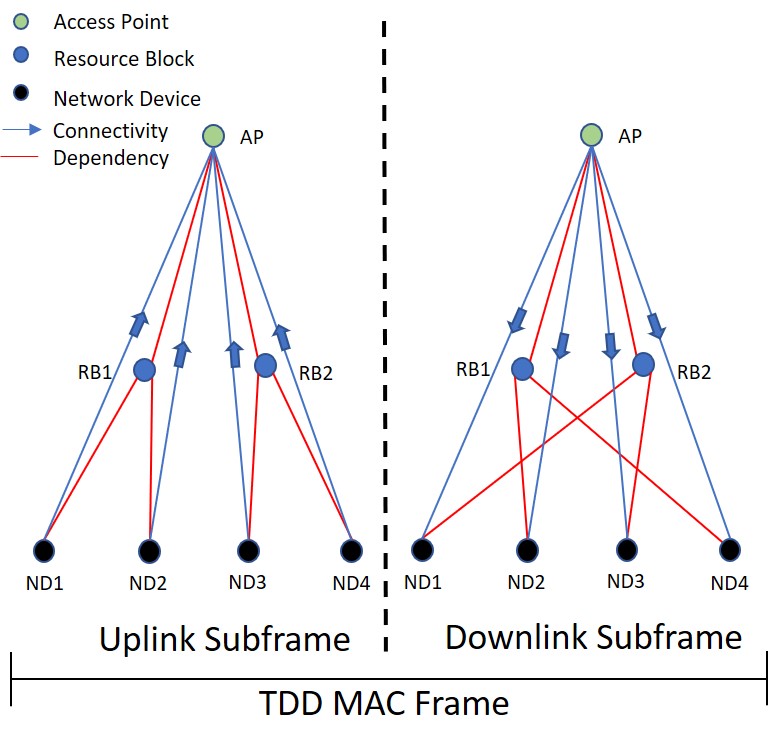}
	\caption{Dependency-Connectivity NOMA Graph Model.}
	\label{fig: NOMAmodel}
\end{figure}

The relationship between NDs, RBs, and the access point (AP) in a NOMA wireless network can be represented by a graph with a combination of dependency and functional connectivity edges as depicted in Figure \ref{fig: NOMAmodel}.
The two graphs correspond to the time division duplex (TDD) uplink and downlink subframes of a single MAC frame. The NDs are dependent on the use of RBs to transmit (receive) information to (from) the AP. The dependency links show the use of a specific RB by a ND and the directed connectivity links show the transmission (reception) of information. The downlink subframe only contains directed connectivity links from the AP to the NDs and the uplink subframe only contains directed connectivity links from the NDs to the AP. The subframe graphs are individual realizations of the network dependency and connectivity links, and the TDD MAC frame represents the smallest time unit in which two different NDs can be bidirectionally connected. 

In the context of 5G NR, this model directly corresponds to two consecutive slots with 15 kHz subcarrier spacing (i.e., the subframe and slot length are equal) in which all symbols are configured for uplink in the first slot and downlink in the second. However, the frame and subframe are general model constructs and can be adapted to any 5G NR subcarrier numerology.

%In the following sections, we refine this general model into a temporal network ensemble parameterized by the number of NDs, RBs, and the overloading ratio. 

\subsection{Temporal network ensemble} \label{netens}
The critical aspect in our model is the dependency relationship between the NDs and the RBs. Given that all the RBs are mapped to a single AP and the directed connectivity can be inferred from the subframe in which the dependency links occur, the mixed dependency-connectivity graphs in Figure \ref{fig: NOMAmodel} can be reduced to a bipartite graph $G(U,V,E)$ where $U$ is the node set of RBs, $V$ is the node set of NDs, and $E$ is the edge set representing the allocation of RBs to NDs in each subframe. 

We assume all RBs are allocated to NDs in each subframe and let the overloading ratio $z \in \mathbb{Z}^+$ and quantity of RBs $|U| \in \mathbb{Z}^+$ both be positive integers. Further, we let $|V|>z|U|$, $\text{deg}\ u=z,\ \forall u \in U$, and $\text{deg}\ v \in \{0,1\},\ \forall v \in V$. This representation allows each ND to potentially form an edge with no more than a single RB, and models the overloading ratio in the degree of each RB node. Since the degree of all RB nodes is the same and the degree of all ND nodes is either one or zero, the network can be specified by a labeled binary degree sequence $\mathbf{d}_v$ (degree vector) that indicates which NDs have received a RB allocation. Thus, the ensemble of possible network configurations is defined by the ensemble of all unique degree vectors of the ND node set parameterized by the number of NDs, RBs, and the overloading ratio $z$. We randomly and uniformly select a degree vector from the ensemble with replacement to generate a graph realization for a NOMA wireless network subframe. This sampling process is repeated until a temporal network with the required number of subframes is generated. The quantity of degree vectors that compose the temporal network ensemble is given by the binomial coefficient, 
\begin{equation} \label{eq1}
\binom{m}{zn} = \frac{m!}{zn!(m-zn)!}, 
\end{equation}
where $m=|V|$ is the number of NDs, $n=|U|$ is the number of RBs, and $z$ is the overloading ratio. Additionally, we note that setting the maximum degree of ND nodes equal to one abstracts the implementation of some code-domain NOMA approaches, but still characterizes the aggregate allocation of RBs to NDs.  

\subsection{Relationship to existing graph models} \label{existingmodels}
Our proposed NOMA network representation is a hybrid adaptation of the well-known configuration model for static networks \cite{bollobas1980probabilistic}, and activity-driven model for temporal networks \cite{perra2012activity}. 

The configuration model is a random graph model that generates an ensemble of graphs from a set of nodes and a given degree sequence. The degree sequence is instantiated in the nodes by assigning a number of ``stubs” equal to the degree of each node in the degree sequence. Two stubs are then chosen uniformly at random to connect and form an edge. The network ensemble is comprised of all possible stub ``matchings," or the different ways in which these stubs can be connected \cite{bianconi2018multilayer,newman2018networks,bollobas1980probabilistic}. However, this ensemble does not suit our purpose due to the unique nature of the explicit dependency and implied connectivity model. Once the stubs are assigned to NDs and RBs, we understand which NDs will have connectivity during that subframe. The number of different ways in which the ND stubs can be mapped to the RB stubs does not provide any additional information about the connectivity. Our model considers whether NDs can map to \emph{any} RB by defining the ensemble as all possible matchings of stubs to nodes rather than all possible ways of connecting stubs that have already been assigned to nodes. 

The activity-driven model is a snapshot representation of a temporal network that assigns an activity potential drawn from a probability distribution to each node in each time step. The activity potential defines a probability that a node forms $m$ undirected edges with $m$ uniformly and randomly selected nodes \cite{perra2012activity,holme2015modern,masuda2020guide}. Nodes which are inactive in a snapshot may still form an edge with another node if they are randomly selected for edge creation by an active node. The probability of RB allocation to each ND (discussed in Section \ref{netseqBP}) is similar to the activity potential, but remains static in each snapshot since it is defined by the NOMA network parameters. Additionally, each ND node that receives a RB allocation can form an edge with any RB node, but this is a dependency relationship. All functional connectivity edges terminate with the AP.

\section{Stochastic temporal component framework} \label{framework}
In this section, we develop a stochastic framework to evaluate temporal component membership. %This analysis provides a measure of robustness for IoT applications with periodic reporting constraints \cite{park2020centralized}.

\subsection{Network sequence as a Bernoulli process} \label{netseqBP}
\begin{figure}[h]
    \centering
	\includegraphics[scale=0.5]{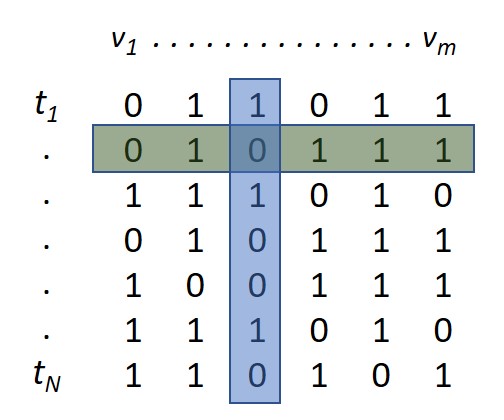}
	\caption{Network Sequence Matrix}
	\label{fig: BernoulliProcess}      
\end{figure}

Allocating RBs to any individual ND in a graph realization generated from a degree vector drawn from the temporal network ensemble can be modeled as a Bernoulli random variable $K$ with probability mass function (PMF) given by
\begin{equation} \label{eq2}
\text{f}_K[k]=
\begin{cases}
p & k=1\\ 1-p & k=0\\ 0 & \text{otherwise}
\end{cases},
\end{equation}
where
\begin{equation} \label{eq3}
p=\frac{zn}{m}.
\end{equation}
As discussed in Section \ref{netens}, the temporal NOMA wireless network is generated from random uniform sampling of the network ensemble. This means the RB allocation is an independent Bernoulli trial in each graph realization for a single ND. Thus, a sequence of $N$ binary degree vectors randomly and uniformly sampled from the network ensemble constitute a Bernoulli random process with $N$ realizations occurring across all labeled nodes $v \in V$ with a probability mass function for each realization given by Equation~\ref{eq2}. In matrix form, this is an $N \times\ m$ matrix, where the rows are the randomly selected degree vectors in ascending temporal order from $1$ to $N$, the columns are the labeled nodes of $V$, and the $ij^{th}$ elements are the degree of node $v_j$ in network realization $t_i$. Thus, each column represents the Bernoulli random process of each labeled node. A visual representation is shown in Figure \ref{fig: BernoulliProcess}. The green rectangle highlights a single network realization occurring at $t_i$, and the blue rectangle highlights a Bernoulli process of RB allocation for a single node, $v_j$. 

\subsection{Probability of resource block allocation} \label{probRBtime}
Characterizing the network sequence of each ND as a Bernoulli random process provides a basis to consider the probability that a ND receives $k$ RB allocations in a network sequence of $N$ realizations. The statistical independence of each RB allocation event means the joint probability of any specific RB allocation sequence is the product all independent probabilities. If the specific order of the RB allocations is not relevant, then we also multiply the joint probability by the number of combinations that could result from a specific number of RB allocations and ``nulls." This combined product is the well-known binomial distribution. Thus, the probability of $k$ RB allocations in a network sequence of $N$ realizations is a binomial random variable with probability mass function
\begin{equation} \label{eq4}
\text{f}_K[k]=
\begin{cases}
\binom{N}{k}p^k(1-p)^{N-k} & 0 \leq k \leq N\\
0 & \text{otherwise}
\end{cases}
\end{equation}
where $p$ is given by Equation~\ref{eq3}. This distribution can also be used to determine the probability that the number of RB allocations falls within a range that is bounded by network sequence length, and can be approximated by a Gaussian distribution as $N$ becomes large ~\cite{papoulis2002prob}. 

Though our RB allocation model is a departure from channel-based dynamic scheduling, the Bernoulli random process ensures each ND receives a nearly equal number of RB allocations as $N$ becomes large \cite{papoulis2002prob}. Thus, our approach provides a maximum fairness baseline for RB allocation that is applicable to narrow band transmissions in which minimum data rate requirements are met for all NDs \cite{shahab2020grant}. 

\subsection{Temporal component membership} \label{tempmem}
Network device membership in a temporal component can be defined in terms of connectivity with the AP, or connectivity with the other NDs in the network. We focus on the latter definition, and treat the AP as a relay rather than a gateway as might be the case in 5G AANs \cite{3gpp.38.811}. This requires a modification to the definition of weak connectedness in \cite{nicosia2012components,nicosia2013graph} because the network topology renders the TWCC indistinguishable from the TSCC. Thus, we adopt the temporal path definition of \cite{nicosia2012components,nicosia2013graph}, but amend the definition of weak temporal connectedness as follows: 

\emph{Weak Temporal Connectedness}: two nodes $i$ and $j$ of a time-varying graph are weakly connected if, either $i$ is temporally connected to $j$, \underline{\emph{or}} $j$ is temporally connected to $i$, in the underlying undirected time-varying graph.

The TWCC definition remains as stated in Section~\ref{temp}.

\section{Individual probability of temporal component membership} \label{indprobtempmem}
\begin{figure}[!ht]
    \centering
    \resizebox{225pt}{125pt}{%}
    \begin{tikzpicture}[grow=right, sloped]
    \node[bag] {Subframe~1 $t_0$}
        child {
            node[bag] {Subframe~2 $t_1$}        
                child {
                    node[end, label=right:
                        {\text{Pr[00]} $=(1-p)^2$}] {}
                    edge from parent
                    node[above] {0}
                    node[below]  {$(1-p)$}
                }
                child {
                    node[end, label=right:
                        {\text{Pr[01]} $=p(1-p)$}] {}
                    edge from parent
                    node[above] {1}
                    node[below]  {$p$}
                }
                edge from parent 
                node[above] {0}
                node[below]  {$1-p$}
        }
        child {
            node[bag] {Subframe~2 $t_1$}        
                child {
                    node[end, label=right:
                        {\text{Pr[10]} $=p(1-p)$}] {}
                    edge from parent
                    node[above] {0}
                    node[below]  {$1-p$}
                }
                child {
                    node[end, label=right:
                        {\text{Pr[11]} $=p^2$}] {}
                    edge from parent
                    node[above] {1}
                    node[below]  {$p$}
                }
            edge from parent         
                node[above] {1}
                node[below]  {$p$}
        };
    \end{tikzpicture}
    }
  \caption{Temporal Component Event Tree}
  \label{fig:eventtree}
\end{figure}
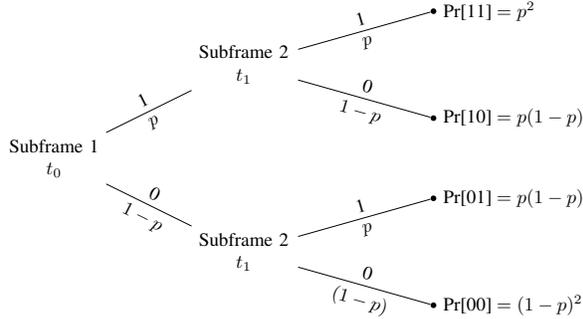

Defining temporal component membership based on connectivity between NDs through the AP requires two subframes to evaluate which NDs belong to which temporal components. We assume all NOMA wireless networks of interest are parameterized by a sufficient combination of RBs and overloading ratio to support RB allocation to more than one ND in a subframe, and correspondingly define the temporal component event tree in Figure \ref{fig:eventtree}. 

This event tree shows the temporal RB allocation sample space for a single ND over two subframes, $t_0$ and $t_1$ (uplink and downlink). Resource block allocation is designated by the value 1 with probability $p$, and a lack of RB allocation (null) is designated by a 0 with probability $1-p$. Two subframes yield a binary sequence of length two with four possible outcomes, $\{11,10,01,00\}$. Strong temporal connectedness requires bidirectional temporal paths between two NDs in the directed graph, which corresponds to a RB allocation in both subframes, e.g.~$\{11\}$. Weak temporal connectedness requires a unidirectional temporal path between two NDs in the underlying undirected time-varying graph which corresponds to one RB allocation in either subframe, either $\{10\}$ or $\{01\}$. Finally, an isolated node corresponds to zero RB allocations in either subframe, $\{00\}$. These mutually exclusive and exhaustive event probabilities are the result of an underlying independent identically distributed (IID) random process; thus, we use each event probability and its complement to define a new Bernoulli random process that characterizes the individual ND membership probability for the corresponding temporal component.

Let the probability that any ND $u \in U$ is a member of the TSCC, TWCC (only), or isolated during any frame in a network sequence be denoted by $p_s$, $p_w$, and $p_i$, respectively, where
\begin{equation*}
    p_s=p^2,\ p_w=2p(1-p),\ p_i=(1-p)^2,\ \text{and}
\end{equation*}
\begin{equation*}
    p_s+p_w+p_i=1.
\end{equation*} 
The probability that a ND is a member of the TSCC in any arbitrary frame is given by substituting $p_s$ for $p$ in Equation \ref{eq2}. Similarly, the probability that a ND is a member of the TSCC for $k$ frames in a network sequence of $N$ frames is given by substituting $p_s$ for $p$ in Equation \ref{eq4}. This same approach also applies to the event probabilities for the TWCC and isolated events. 

\section{Joint probability of temporal component membership} \label{jointprobtempmem}
The analysis of joint ND temporal component membership is different than for an individual ND. Recall from Section \ref{netseqBP} that we can analyze one ND over an arbitrary number of network realizations from $\{1,...,N\}$, where each realization is an IID Bernoulli random variable, and the joint probability of any specific sequence of events is the product of the individual event probabilities. However, each network realization is the result of a degree vector with fixed values dependent on the network parameters, so the probability of RB allocation for two or more NDs are not independent within a \emph{single} network realization. 

The binary characterization of RB allocation facilitates use of the Hypergeometric distribution for joint RB probability analysis within a single network realization. The distribution defines the probability of drawing a sample of $n$ items containing $k$ successes without replacement from a population $N$ with a known number of successes $M$ \cite{papoulis2002prob}. The PMF is given by 
\begin{equation} \label{eq5}
    \text{f}_K[k]=\frac{\binom{M}{k} \binom{N-M}{n-k}}{\binom{N}{n}}.
\end{equation} 
In our NOMA network model, the successes are those NDs in the sample that receive a RB allocation. Substituting the NOMA parameters into Equation \ref{eq5} gives
\begin{equation} \label{eq6}
    \text{f}_K[k]=\frac{\binom{nz}{k} \binom{m-nz}{m_s-k}}{\binom{m}{m_s}},
\end{equation}
where $m_s$ is the number of NDs sampled, $m$ is the total number of NDs in the network, the quantity of known successes is the number of RBs $n$ multiplied by the overloading ratio $z$, and $k$ is the number of RBs allocated to the unlabeled NDs in the sample. 

Determining the joint probability of RB allocation among any \emph{labeled} combination of NDs requires equally distributing the unlabeled joint probability over the number of possible unique RB allocations to the labeled NDs, where the number of possible RB combinations is given by the binomial coefficient $\binom{m_s}{k}$. Thus, the labeled probabilities are given by
\begin{equation} \label{eq7}
    \text{f}_K[k]_{\text{labeled}}=\frac{\text{f}_K[k]}{\binom{m_s}{k}}.
\end{equation}

Once the joint probability has been determined, it is fixed over all frames in the network sequence and each frame is an IID realization. Thus, we can define an exhaustive and mutually exclusive joint probability space with the desired number of NDs and calculate temporal component membership over an arbitrary number of network realizations using the same approach discussed in Section \ref{indprobtempmem}. 

\section{Affine graph analysis}
\label{affineanalysis}
In this section, we consider the minimum number of MAC frames required to achieve a complete affine graph and the probability of occurrence. Recall that the affine graph provides a temporal measure of bidirectional connectivity between ND pairs in the network. Pairwise ND communication is central in distributed computing; thus, this analysis relates NOMA overloading to the supported convergence time of distributed computing algorithms that could be deployed on 5G AANs \cite{ nedic2018network,shi2020communication}.

Despite research that shows more efficient wireless communications methods using downlink broadcast for distributed computation \cite{li2017scalable}, our analysis considers individual scheduling for NDs in the downlink subframe. This is consistent with the 5G NR MAC sublayer logical to transport channel mapping. All user-plane traffic is mapped from the dedicated traffic channel (logical) to the downlink shared channel (transport), and not to a broadcast channel which is reserved for system information \cite{dahlman20185g,ahmadi20195g}. %Including user-plane broadcast at the MAC sublayer would require a cross-layer modification to the NR standard. in which the broadcast requirement from the distributed computing application was signaled in a way that that resulted in a mapping to a broadcast transport channel. 

\subsection{Mathematical framework} \label{mathframe}
Our mathematical framework for affine graph analysis includes connectivity potential matrices, a frame connectivity matrix, and a queue matrix. 

Connectivity potential matrices translate the allocation of RBs into the corresponding potential connectivity they enable between NDs.

\underline{\emph{Uplink Potential Matrix}}: Given a binary RB allocation vector $\mathbf{d}_v=[d_1,...,d_m]$ for $m$ NDs, let $\mathbf{P}_{UL}$ be the uplink potential matrix given by
\begin{equation} \label{eq8}
    \mathbf{P}_{UL}=(\mathbf{1} - \mathbf{I}) \odot \mathbf{d}^T,
\end{equation}
where $\mathbf{1}$ is $m \times m$ uniform matrix, $\mathbf{I}$ is $m \times m$ identity matrix, and $\odot$ denotes the Hadamard Product.

\underline{\emph{Downlink Potential Matrix}}: Given a binary RB allocation vector $\mathbf{d}_v=[d_1,...,d_m]$ for $m$ NDs, let $\mathbf{P}_{DL}$ be the downlink potential matrix given by
\begin{equation} \label{eq9}
    \mathbf{P}_{DL}=(\mathbf{1} - \mathbf{I}) \odot \mathbf{d}.
\end{equation}

\underline{\emph{Frame Connectivity Matrix}}: The frame connectivity matrix $\mathbf{F}$ is a directed adjacency matrix that defines the connectivity achieved between NDs during a single frame as a result of the uplink and downlink RB allocations. The frame connectivity matrix is the Hadamard Product, of the uplink and downlink potential matrices, $\mathbf{P}_{UL}$ and $\mathbf{P}_{DL}$, given by 
\begin{equation} \label{eq10}
    \mathbf{F} = \mathbf{P}_{UL} \odot \mathbf{P}_{DL}.
\end{equation}

\underline{\emph{Frame Queue Matrix}}: The frame connectivity matrix defines the achieved connectivity in each frame, but does not account for ``undelivered" messages that result from the $ij$th elements of the uplink potential matrix that go unmatched with the downlink potential matrix in the same frame. This information must be captured since a pair of NDs that were unable to communicate over the time interval of one frame might be able to communicate if the time interval is extended to two frames. A queue matrix to capture this information corresponds to buffers in wireless networks where packets are queued until they can be transmitted. This matrix will change during each frame depending on the uplink and downlink RB allocations, so we define the queue matrix $\mathbf{Q}$ recursively as
\begin{equation} \label{eq11}
    \mathbf{Q}_{t} = \mathbf{P}_{UL}+\mathbf{Q}_{t-1}-\mathbf{F}_t,
\end{equation}
where $\mathbf{Q}_{0}$ is the $m \times m$ null matrix $\mathbf{0}$. The queue matrix that feeds into the next frame is the sum of the uplink potential matrix and previous queue matrix, minus the achieved transmissions resulting from the frame connectivity matrix. 

\underline{\emph{Dynamic Frame Connectivity Matrix}}: The time- dependent definition of the frame queue matrix $\mathbf{Q}_t$ implies the frame connectivity matrix $\mathbf{F}$ must account for queued messages from the previous frame. We define the frame connectivity matrix at time $t$ 
\begin{equation} \label{eq12}
    \mathbf{F}_t = \mathbf{1}_{\mathbb{Z}^+}(\mathbf{P}_{UL}+\mathbf{Q}_{t-1}) \odot \mathbf{P}_{DL},
\end{equation}
where $\mathbf{1}_{\mathbb{Z}^+}(\mathbf{X})$ is the indicator function for the set of positive integers to generate a binary matrix. This conceptualization captures the directed connectivity between nodes that occurs over more than one frame.

\subsection{Affine graph} \label{affgraphcon}
Each dynamic frame connectivity matrix captures the achieved connectivity in frame $t$. However, the affine graph represents strong temporal connectedness between node pairs over an observation interval. Thus, we must sum all the individual frame connectivity matrices to produce an aggregate picture of network connectivity during an arbitrary observation interval $t=[T-l,T-l+1,...,T-1,T]$ selected from the network sequence. If we let $l=T+1$, then $t$ is the entire network sequence, and the static projection directed adjacency matrix of the NOMA wireless network is given by
\begin{equation} \label{eq13}
    \mathbf{S}=\sum_{t=1}^T \mathbf{F}_t.
\end{equation}
Since $\mathbf{S}$ represents all achieved connectivity over the entire network sequence, we calculate the affine graph 
\begin{equation} \label{eq14}
\mathbf{A} = \mathbf{1}_{\mathbb{Z}^+}(\mathbf{S} \odot \mathbf{S}^T)
\end{equation}
Recall that the $i$th row of $\mathbf{S}$ represents the achieved directional connectivity from ND $i$ to ND $j$ ($\forall j \; \text{with}\ i \neq j$), and the $j$th column of $\mathbf{S}$ represents the achieved directional connectivity from ND $j$ to ND $i$. Thus, taking the Hadamard Product of $\mathbf{S}$ with its transpose only yields a non-zero value for the $ij$th element of $\mathbf{A}$ if bidirectional connectivity has been established between nodes $i$ and $j$. Applying the indicator function results in an unweighted binary matrix. 

\subsection{Analytical results} \label{afftimedist}
The minimum number of frames required to form a complete affine graph is the sum of the number of frames required for all NDs to receive an uplink RB allocation, and the number of frames required for all NDs to receive a downlink RB allocation after all have received an uplink RB allocation. The corresponding probability of this event depends on the network parameters $m$, $n$, and $z$. We now present the possible cases for this relationship. 

\underline{\emph{Case 1}}: $zn = m-1$. This case occurs when all but one ND can receive a RB allocation in each subframe. A complete affine graph can be achieved in two frames with probability 
\begin{equation} \label{eq15}
    \Pr[F_{min}]=p^2(1-p),
\end{equation}
where $p$ is the probability of RB allocation for an individual ND defined in Equation \ref{eq3}. 

Consider the case in which the single ND $i$ that does not receive an uplink RB allocation in the first subframe does receive a RB allocation in the first downlink subframe. This means ND $i$ has received from all other NDs (since ND $i$ does not transmit to itself). If ND $i$ receives an uplink RB allocation in the second uplink subframe, then all NDs have received an uplink RB allocation. Since ND $i$ has already received from all other NDs, it does not require an additional downlink RB allocation, so if all other $m-1$ NDs receive a downlink RB allocation in the second downlink subframe, then a complete affine graph will have been achieved. This sequence of frames can be characterized entirely by the behavior of a single ND $i$, with probability of RB allocation, $p$.

\underline{\emph{Case 2}}: $\frac{m}{2} \leq zn < m-1$. This case occurs when at least half of the NDs in the network can receive a RB allocation in each subframe, but less than $\text{ND}-1$. A complete affine graph can be achieved in three frames with probability 
\begin{equation} \label{eq16}
    \Pr[F_{min}]=\left[\frac{\binom{zn}{2zn-m}}{\binom{m}{zn}}\right]^2.
\end{equation}
In this case, a minimum of two uplink subframes are required for all NDs to receive at least one uplink RB allocation. Similarly, a minimum of two downlink subframes are required for all NDs to receive at least one downlink RB allocation. All NDs must receive a downlink RB allocation \emph{after} all NDs have received an uplink RB allocation in order to receive a transmission directly in the same frame, or indirectly from the message queue. Thus, the downlink RB allocation must begin in the second downlink subframe, and conclude in the third downlink subframe.

Consider a NOMA wireless network parameterized by $m=6$, $n=2$, and $z=2$. In the first frame, the probability of selecting four NDs which have not yet received an uplink RB allocation is equal to one since the network sequence has just begun. In the second frame, we must select the $m-zn=2$ NDs that did not receive an uplink RB allocation in the first uplink subframe, as well as $zn-(m-zn)=2$ of the $zn=4$ NDs that did receive an uplink RB allocation in the first uplink subframe. The probability of this event is \[\frac{\binom{m-zn}{m-zn}\binom{zn}{zn-(m-zn)}}{\binom{m}{zn}}=\frac{\binom{zn}{2zn-m}}{\binom{m}{zn}}=\frac{\binom{2}{2}\binom{4}{2}}{\binom{6}{4}}=\frac{\binom{4}{2}}{\binom{6}{4}}.\] The second downlink subframe is the first opportunity for NDs to receive a downlink RB allocation after all NDs have received an uplink RB allocation. Thus, the RB allocation sequence that was required for the uplink in the first and second frames must repeat for the downlink in the second and third frames to achieve a complete affine graph, which gives \[\Pr[F_{min}]=\left[\frac{\binom{4}{2}}{\binom{6}{4}}\right]^2.\]
Generalizing this to any combination of NDs, RBs, and overloading ratio that meet these conditions results in Equation \ref{eq16}. Note that when $zn = \frac{m}{2}$, Equation \ref{eq16} reduces to 
\begin{equation} \label{eq17}
    \Pr[F_{min}]=\binom{m}{m/2}^{-2}.
\end{equation}

\underline{\emph{Case 3}}: $1 < zn < \frac{m}{2}$. This case occurs when more than one, but less than  half, of the NDs can receive a RB allocation in each subframe. There are three sub-cases within Case 3. 

\emph{Subcase 3.1}: $m \mod zn = 0$. This occurs when the product of RBs and the overloading ratio divide evenly into the number of NDs. The number of frames required for all NDs to receive a RB allocation is equal to $\frac{ND}{zn}$, and the downlink RB allocation can begin in the same frame that all NDs receive an uplink allocation. Hence, the minimum number of frames required to achieve a complete affine graph is given by 
\begin{equation} \label{eq18}
    F_{min}=\frac{2m}{zn}-1.
\end{equation}
The probability of this event is a product of the Hypergeometric distribution evaluated with constant population ($m$), sample size ($zn$), and desired number of successes ($zn$), but a decreasing number of known successes in the population. We treat the NDs that have yet to receive an uplink or downlink RB allocation as the number of successes in the population, and decrement this value by $zn$ after each subframe. The probability of selecting $zn$ NDs that have not yet received a RB allocation in the first uplink subframe is one since no NDs have received an RB allocation at the beginning of the network sequence. In the second uplink subframe, we must select $zn$ of $m-zn$ NDs that have not yet received an uplink RB allocation. This decrementing operation continues until all NDs have received an uplink RB allocation, and then repeats for the downlink, resulting in a probability given by
\begin{equation} \label{eq19}
    \Pr[F_{min}]=\prod_{i=1}^{\frac{m}{zn}-1}\frac{\binom{m-i(zn)}{zn}^2}{\binom{m}{zn}^2}.
\end{equation}
Since we never select any NDs that have already received a RB allocation, the second term in the numerator from Equation \ref{eq6} reduces to one and the square operation accounts for the repetition of the process in the uplink and downlink.

\emph{Subcase 3.2}: $m \mod zn = 1$. This occurs when any multiple of the product of RBs and the overloading ratio is one less than the number of NDs. The logic for the minimum frames required for a complete affine graph is the same as Case 1, except there are more frames prior to reaching the frame in which only a single ND has not yet received an uplink RB allocation. Accounting for those additional frames, the minimum number of frames required to achieve a complete affine graph is given by 
\begin{equation} \label{eq20}
    F_{min}=2\bigg \lfloor \frac{m}{zn} \bigg \rfloor. 
\end{equation}
The probability formulation for this case is similar to Subcase 3.1 in that we take a product of Hypergeometric distributions evaluated at decrementing values of the number of successes in the population. The primary differences occur in  downlink subframe $\big \lfloor \frac{m}{zn} \big \rfloor$ and in uplink subframe $\big \lfloor \frac{m}{zn} \big \rfloor+1$ which immediately follows. In  downlink subframe $\big \lfloor \frac{m}{zn} \big \rfloor$, all NDs have received an uplink RB allocation except ND $i$. Thus, ND $i$ must receive a downlink RB allocation in this subframe to have received from all other NDs. This occurs with probability $\binom{m-1}{zn-1}/\binom{m}{zn}$. Similarly, ND $i$ must receive an uplink RB allocation in the following uplink subframe so all remaining $m-1$ NDs can receive from ND $i$ in the following $\big \lfloor \frac{m}{zn} \big \rfloor$ subframes. This also occurs occurs with probability $\binom{m-1}{zn-1}/\binom{m}{zn}$. Thus, the total probability of achieving a complete affine graph in this subcase is given by
\begin{equation} \label{eq21}
\begin{split} 
    \Pr[F_{min}]=\frac{\binom{m-1}{zn-1}^2}{\binom{m}{zn}^{2\alpha+1}} \prod_{i=1}^{\alpha-1}\binom{m-izn}{zn} \\ 
    \prod_{j=0}^{\alpha-1}\binom{m-1-jzn}{zn},
\end{split}
\end{equation}
where we have collected all denominators of the product into the denominator of the first term, the second and third terms are the decrementing Hypergeometric distributions for the uplink and downlink, respectively, and $\alpha=\Big \lfloor \frac{m}{zn} \Big \rfloor$.

\emph{Subcase 3.3}: $m \mod zn > 1$. This occurs when any multiple of the product of RBs and the overloading ratio does not fall in either Subcase 3.1 or 3.2. The minimum number of frames required to achieve a complete affine graph is given by 
\begin{equation} \label{eq22}
    F_{min}=2\bigg \lceil \frac{m}{zn} \bigg \rceil-1.
\end{equation}
The probability of achieving the affine graph in the minimum number of frames can be directly calculated, but does not have a concise expression. Rather, we must determine all possible ways in which all NDs can receive a RB allocation in the minimum number of frames, determine the probability of each of these possibilities, and then sum the pairwise multiplication of all probabilities to account for all possible uplink and downlink combinations of RB allocations. 

\subsection{Simulation results}
Simulations to test our analytical results were conducted in MATLAB. For each case with an expression for $\Pr[F_{min}]$ we defined a number of NDs ($m$), RBs ($n$), and overloading ratio ($z$) that met the case conditions and generated the corresponding temporal network ensemble. Next, we generated a network sequence of length $N$ through random uniform sampling from the ensemble where $N$ was set equal to $F_{min}$ based on the appropriate case. Finally, we calculated the resulting affine graph from each network sequence using the mathematical framework of Sections \ref{mathframe} and \ref{affgraphcon}. 

We generated $5\times 10^6$ network sequences for each case and recorded the number of complete affine graphs achieved. We compared the simulation result with the predicted value for each case and calculated the absolute percent error. The cases, network parameters, and error results are shown in Table \ref{tab: results}.
\begin{table}[thb]
\centering
\caption{Simulation Results}
\label{tab: results}
\begin{tabular}{ |c| c c c |c| }
 \hline
 {Case} &ND &RB &$z$ &Error (\%)\\
 \hline
 1 &9 &2 &4 &0.1341\\
 2 &10 &2 &4 &0.0574 \\
 2 &10 &2 &2.5 &0.9338 \\
 3.1 &6 &2 &1 &0.2125 \\
 3.2 &9 &2 &2 &0.1117 \\
\hline
\end{tabular}
\end{table}

The largest absolute percent error across all cases is 0.9338\% which lends strong support for the accuracy of the analytical results. We note this error corresponds to the lower bound of Case 2 in which exactly half of the NDs can receive a RB allocation in each frame. This relationship between $m$ and $zn$ corresponds to the maximum size of the network ensemble described in Equation \ref{eq1}, so we expect a larger error than other simulation cases with a smaller network ensemble.  

\section{Conclusion} \label{conc}

In this paper, we explored the relationship between  NOMA overloading and network robustness through temporal connectedness in the context of a 5G AAN. We proposed a mixed dependency-connectivity graph model for a NOMA network, developed a stochastic temporal component framework, and characterized the probability of temporal component membership in terms of the overloading ratio. We extended this analysis to the affine graph by developing analytical expressions for the minimum number of MAC frames in which bidirectional connectivity can be achieved between all NDs, and the probability of these events. Finally, we supported our analytical results through simulation. 

%Bibliography 
\bibliographystyle{ieeetr}
\bibliography{final}

\end{document}